\documentclass[vecphys]{svmult}
\usepackage{makeidx}         
\usepackage{graphicx}        
\usepackage{multicol}        
\usepackage[bottom]{footmisc}
\makeindex             

\begin{document}

\title*{Binary Stars as the Source of the Far-UV Excess
in Elliptical Galaxies}

\titlerunning{Far-UV Excess in Elliptical Galaxies}

\author{Zhanwen Han\inst{1}, Philipp Podsiadlowski\inst{2}, 
Anthony E. Lynas-Gray\inst{2}}
\institute{National Astronomical Observatories / Yunnan Observatory,
the Chinese Academy of Sciences, Kunming, 650011, China\\
\texttt{zhanwenhan@hotmail.com}
\and
University of Oxford, Department of Physics,
Oxford, OX1 3RH, UK \\
\texttt{podsi@astro.ox.ac.uk, aelg@astro.ox.ac.uk}}

\maketitle

\begin{abstract}
 The discovery of an excess of light in the far-ultraviolet (UV)
 spectrum in elliptical galaxies was a major surprise in 1969.  While
 it is now clear that this UV excess is caused by an old population of
 hot helium-burning stars without large hydrogen-rich envelopes rather
 than young stars, their origin has remained a mystery.  Here we show
 that these stars most likely lost their envelopes because of binary
 interactions, similar to the hot subdwarf population in our own
 Galaxy. This has major implications for understanding the evolution
 of the UV excess and of elliptical galaxies in general. In
 particular, it implies that the UV excess is not a sign of age, as
 had been postulated previously, and predicts that it should not be
 strongly dependent on the metallicity of the population.
\end{abstract}

\section{Introduction}

One of the first major discoveries soon after the advent of UV
astronomy was the discovery of an excess of light in the
far-ultraviolet (far-UV) in elliptical galaxies (see the review by
O'Connell \cite{oco99}). This came as a complete surprise since
elliptical galaxies were supposed to be entirely composed of old, red
stars and not to contain any young stars that radiate in the UV.
Since then it has become clear that the far-UV excess (or upturn) is
not a sign of active contemporary star formation, but is caused by an
older population of helium-burning stars or their descendants with a
characteristic surface temperature of 25,000\,K \cite{fer91}, also known
as hot subdwarfs.

The origin of this population of hot, blue stars in an otherwise red
population has, however, remained a major mystery \cite{gre90}.  Two
scenarios, referred to as the high- and the low-metallicity scenario,
have been advanced. In the low-metallicity model \cite{lee94}, it is
argued that these hot subdwarfs originate from a low-metallicity
population of stars which produce very blue helium core-burning
stars. This model tends to require a very large age of the population
(in fact, larger than the generally accepted age of the Universe); it
is also not clear whether the population is sufficiently blue to
account for the observed UV color. Moreover, the required low
metallicity appears to be inconsistent with the large metallicity
inferred for the majority of stars in elliptical galaxies \cite{ter02}. 
In contrast, the high-metallicity model \cite{bre94,yi97} assumes a 
relatively high metallicity -- consistent with the metallicity of
typical elliptical galaxies ($\sim 1$\,--\,3 times the solar
metallicity) -- and an associated enhancement in the helium abundance
and, most importantly, postulates an enhanced and variable mass-loss
rate on the red-giant branch, so that a fraction of stars lose most of
their hydrogen-rich envelopes before igniting helium in the core 
\cite{yi97,dor95}.

Both models are quite \textit{ad hoc}: there is neither observational
evidence for a very old, low-metallicity sub-population in elliptical
galaxies, nor is there a physical explanation for the very high mass
loss required for just a small subset of stars. Furthermore, both
models require a large age for the hot component and therefore predict
that the UV excess declines rapidly with redshift. This is not
consistent with recent observations, e.g. with the Hubble Space
Telescope (HST) \cite{bro03}.  In particular, the recent survey with the
GALEX satellite \cite{ric05} showed that the UV excess, if anything, may
increase with redshift. Indeed, the wealth of observational data
obtained with GALEX is likely to revolutionize our understanding of
elliptical galaxies.  While Burstein et al.\ \cite{bur88} appeared to
have found a correlation between the UV-upturn and metallicity in
their sample of 24 quiescent elliptical galaxies, which could support
the high-metallicity scenario, this correlation has not been confirmed
in the much larger GALEX sample \cite{ric05}, casting serious doubt on
this scenario.

Both models ignore the effects of binary evolution.  On the other
hand, hot subdwarfs have long been studied in our own Galaxy \cite{heb86}, 
and it is now well established \cite{max01} that the vast majority
of (and quite possibly all) Galactic hot subdwarfs are the results of
binary interactions where a star loses all of its envelope near the
tip of the red-giant branch by mass transfer to a companion star or by
ejecting it in a common-envelope phase, or where two helium white
dwarfs merge to produce a single object (see \cite{han02,han03} for
references and details). In all of these cases, the remnant star
ignites helium and becomes a hot subdwarf.  The key feature of these
binary channels is that they provide the missing physical mechanism
for ejecting the envelope and for producing a hot subdwarf.  Moreover,
since it is known that these hot subdwarfs provide an important source
of far-UV light in our own Galaxy, it is not only reasonable to assume
that they will also contribute significantly to the far-UV in
elliptical galaxies, but is in fact expected.

\section{The Model}

To quantify the importance of the effects of binary interactions
on the spectral appearance of elliptical galaxies, we have
performed the first population synthesis
study of galaxies that includes binary evolution (see also 
\cite{bru93,wor94,zha05}).  It is based on a binary population model
\cite{han02,han03} that has been calibrated to reproduce the short-period
hot subdwarf binaries in our own Galaxy that make up the majority of
Galactic hot subdwarfs \cite{max01}. The population synthesis model
follows the detailed time evolution of both single and binary stars,
including all binary interactions, and is capable of simulating
galaxies of arbitrary complexity, provided the star-formation history
is specified. To obtain galaxy colors and spectra, we have calculated
detailed grids of spectra for hot subdwarfs using the {\scriptsize
ATLAS9} \cite{kur92} stellar atmosphere code.  For the spectra and colors of
single stars with hydrogen-rich envelopes, we use the comprehensive
BaSeL library of theoretical stellar spectra \cite{lej97,lej98}.

\section{Results and Discussion}

\begin{figure}
\centering
\includegraphics[height=9cm,angle=270]{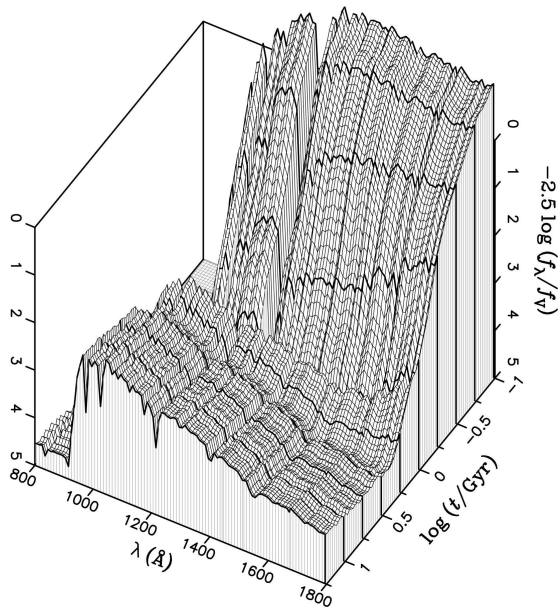}
\caption{The evolution of the far-UV spectrum with time for a single
population where all stars formed at the same time. The flux
$f_{\lambda}$ is scaled relative to the visual flux ($f_{\rm V}$).}
\label{fig1}
\end{figure}

\begin{figure}
\centering
\includegraphics[height=11cm,angle=270]{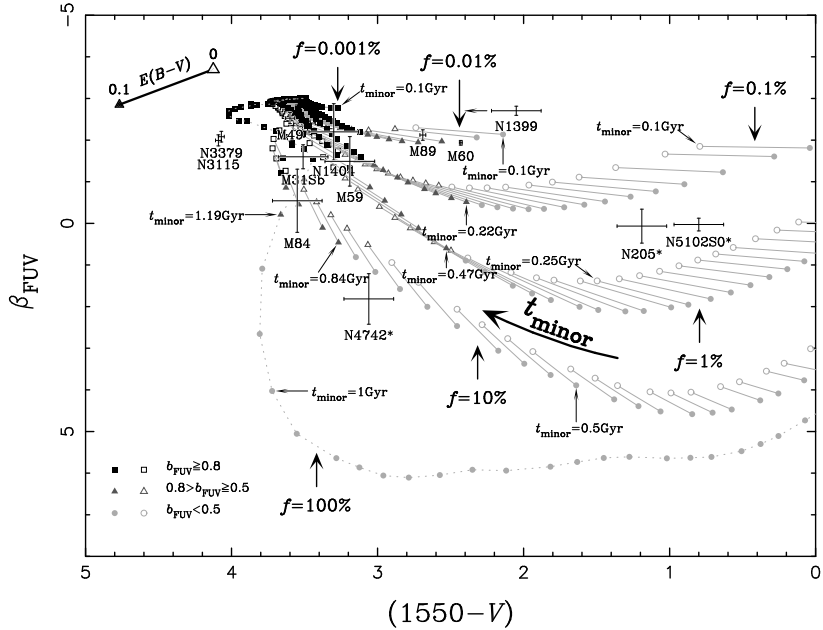}
\caption{
Evolution of far-UV
properties [the slope of the far-UV spectrum, $\beta_{\rm FUV}$,
versus $(1550-V)$] for a two-population model of
elliptical galaxies.
The age of the old population is assumed to be 12\,Gyr (filled
squares, filled triangles, or filled circles) or 5\,Gyr
(open squares, open triangles, or open circles).
The mass fraction of the younger
population is denoted as $f$ and the time since the formation as
$t_{\rm minor}$ [plotted in steps of $\Delta \log (t)=0.025$].
Note that the model for $f=100\%$ (the dotted curve)
shows the evolution of a simple stellar population with age
$t_{\rm minor}$.  The legend is for $b_{\rm FUV}$, which is
the fraction of the UV flux that originates from hot subdwarfs resulting
from binary interactions. The effect of internal extinction is
indicated in the top-left corner, based on the Calzetti internal
extinction model with $E(B-V)=0.1$ \cite{cal00}.
For comparison, we also plot galaxies with error bars from HUT
\cite{bro97} and IUE observations \cite{bur88}.
The galaxies with strong signs of
recent star formation are denoted with an asterisk (NGC 205, NGC 4742,
NGC 5102).
}
\label{fig2}
\end{figure}

Figure~\ref{fig1} shows our simulated evolution of the far-UV spectrum of a
galaxy in which all stars formed at the same time, where the flux has
been scaled relative to the visual flux (between 5000 and 6000\AA) to
reduce the dynamical range. At early times the far-UV flux is
dominated by the contribution from single young stars. Binary hot
subdwarfs become important after about 1.1\,Gyr, which corresponds to
the evolutionary timescale of a 2\,$M_{\odot}$ star and soon start to
dominate completely. After a few Gyr the spectrum no longer changes
appreciably.

There is increasing evidence that many elliptical galaxies had some
recent minor star-formation events \cite{sch06,kav06}, which also
contribute to the far-UV excess.  To model such secondary minor
starbursts, we have constructed two-population galaxy models,
consisting of one old, dominant population with an assumed age $t_{\rm
old}$ and a younger population of variable age, making up a fraction
$f$ of the stellar mass of the system.  
In order to illustrate the appearance of the galaxies for different
lookback times (redshifts), we adopted two values for
$t_{\rm old}$, of 12 Gyr and 5 Gyr, respectively; these values
correspond to the ages of elliptical galaxies at a redshift of 0 and
0.9, respectively, assuming that the initial starburst occurred at a
redshift of 5 and adopting a standard $\Lambda$CDM cosmology with
$H_0=72{\rm km/s/Mpc}$, $\Omega_{\rm M}=0.3$ and
$\Omega_\Lambda=0.7$. Our spectral modelling shows that a recent minor
starburst mostly affects the slope in the far-UV SED. 
We therefore define a far-UV slope index $\beta_{\rm FUV}$ as
$f_\lambda \sim \lambda ^{\beta_{\rm FUV}}$, where $\beta_{\rm FUV}$
is fitted between 1075\AA~ and 1750\AA. This parameter was obtained
from our theoretical models by fitting the far-UV SEDs and was derived
in a similar manner from observed far-UV SEDs of elliptical galaxies
\cite{bur88,bro97}, where we excluded the spectral region between 1175\AA~
and 1250\AA, the region containing the strong Ly$\alpha$ line.  In
order to assess the importance of binary interactions, we also defined
a binary contribution factor $b={F_{\rm b}/ F_{\rm total}}$, where
$F_{\rm b}$ is the integrated flux between 900\AA~ and 1800\AA~
radiated by hot subdwarfs produced by binary interactions, and $F_{\rm
total}$ is the total integrated flux between 900\AA ~and 1800\AA.
Figure~\ref{fig2} shows the far-UV slope as a function of UV excess, a
potentially powerful diagnostic diagram which illustrates how the UV
properties of elliptical galaxies evolve with time in a dominant old
population with a young minor sub-population.  For comparison, we also
plot observed elliptical galaxies from various sources.  Overall, the
model covers the observed range of properties reasonably well.  
Note in particular that the majority of galaxies lie in
the part of the diagram where the UV contribution from binaries is
expected to dominate (i.e. where $b> 0.5$).  

The two-component models presented here are still quite simple and do not
take into account, e.g., more complex star-formation histories, possible
contributions to the UV from AGN activity, non-solar metallicity or
a range of metallicities.  Moreover, the binary
population synthesis is sensitive to uncertainties in the binary
modelling itself, in particular the mass-ratio distribution and the
condition for stable and unstable mass transfer \cite{han03}. We have
varied these parameters and found that these uncertainties do not
change the qualitative picture, but affect some of the quantitative
estimates. 

Despite its simplicity, our model can successfully reproduce most of
the properties of elliptical galaxies with a UV excess: e.g., the
range of observed UV excesses, both in $(1550-V)$ and $(2000-V)$ (e.g.
\cite{deh02}), and their evolution with redshift.  The model predicts
that the UV excess is not a strong function of age, and hence is not a
good indicator for the age of the dominant old population, as has been
argued previously \cite{yi99}, but is very consistent with recent
GALEX findings \cite{ric05}.  We typically find that the $(1550-V)$ color
changes rapidly over the first 1\,Gyr and only varies slowly
thereafter. This also implies that all old galaxies should show a UV
excess at some level. Moreover, we expect that the model is not very
sensitive to the metallicity of the population since metallicity does
not play a significant role in the envelope ejection process (although
it may affect the properties of the binary population in more subtle
ways).

Our model is sensitive to both low levels and high levels of star
formation.  It suggests that elliptical galaxies with the largest UV
excess had some star formation activity in the relatively recent past
($\sim 1\,$Gyr ago).  AGN and supernova activity may provide
supporting evidence for this picture, since the former often appears
to be accompanied by active star formation, while supernovae, both
core collapse and thermonuclear, tend to occur mainly within
1\,--\,2\,Gyr after a starburst in the most favoured supernova models.

The modelling of the UV excess presented in this study is only a
starting point: with refinements in the spectral modelling, including
metallicity effects, and more detailed modelling of the global
evolution of the stellar population in elliptical galaxies, we suspect
that this may become a powerful new tool helping to unravel the
complex histories of elliptical galaxies that a long time ago looked
so simple and straightforward.

This work has been supported by the Chinese National Science Foundation
under Grant Nos.\ 10433030 and 10521001 (ZH). See \cite{han07}
for a detailed version of the paper.



\printindex
\end{document}